\documentstyle[12pt,worldsci]{article}

\def\thefootnote{\fnsymbol{footnote}}
\def\bea {\begin{eqnarray}}
\def\eea {\end{eqnarray}}
\def\be {\begin{equation}}
\def\ee {\end{equation}}
\def\ben{\begin{enumerate}}
\def\een{\end{enumerate}}
\def\bi{\begin{itemize}}
\def\ei{\end{itemize}}
\def\ie{{\it i.e.}}
\def\viz{{\it viz.}\ }

\def\etal{{\it et al.}}

\def\F{{\cal F}}

\def\prl {Phys. Rev. Lett.\ }
\def\pl {Phys. Lett.\ }
\def\pr {Phys. Rev.\ }
\def\np {Nucl. Phys.\ }

\def\GV{G_{\mbox{\tiny V}}}

\def\DRV{\Delta_{\mbox{\tiny R}}^{\mbox{\tiny V}}}

\def\mids{\! \mid \! }

\def\nl{$\,\!$}

\begin{document}

\title{RECENT CHALK RIVER EXPERIMENTS ON SUPERALLOWED $0^{+} \rightarrow
0^{+}$ BETA DECAYS}
\author{I. S. Towner, E. Hagberg, G. Savard, J. C. Hardy, V.T. Koslowsky, \\
A. Galindo-Uribarri and D.C. Radford \\
{\em AECL, Chalk River Laboratories, Chalk River \\
Ontario K0J 1J0, Canada}}
\maketitle

\begin{abstract}
{\footnotesize
Two experiments, (1) a measurement of the superallowed $0^{+}
\rightarrow 0^{+}$ branching ratio in $^{10}$C, and (2) a measurement
of (or limit on) non-analogue $0^{+} \rightarrow 0^{+}$ branches
in $^{38m}$K, $^{46}$V, $^{50}$Mn and $^{54}$Co, are described.
The implications these experiments have on the test of the
conservation of the weak vector current (CVC) and the unitarity
of the Cabibbo-Kobayashi-Maskawa matrix are surveyed.
}

\end{abstract}

\renewcommand{\thefootnote}{\#\arabic{footnote}}
\setcounter{footnote}{0}

\section{Introduction} \label{intro}

Superallowed Fermi $0^+ \rightarrow 0^+$ nuclear beta decays
\cite{Ha90,TH95}
provide both the best test of the Conserved Vector Current (CVC)
hypothesis in weak interactions and, together with the muon
lifetime, the most accurate value for the up-down quark-mixing
matrix element of the Cabibbo-Kobayashi-Maskawa (CKM) matrix,
$V_{ud}$. At present, the deduced value of $V_{ud}$ from nuclear
beta decay is such that, with standard values
\cite{PDG94} of the other elements of the CKM matrix, the
unitarity test from the sum of the squares of the elements in the
first row fails to meet unity by more than twice the estimated
error.

According to CVC, all Fermi decays should yield a
nucleus-independent value of the weak vector coupling constant, $\GV$,
from their measured $ft$ values provided that small
isospin-symmetry-breaking
\cite{THH77,To89,OB89,OB95}
($\delta_C$) and radiative
\cite{Si78,To94}
($\delta_R$) corrections are accounted for.  Specifically for an
isospin-1 multiplet

\be
\F t = ft (1 + \delta_R )(1 - \delta_C) = \frac{K}{2 {\GV^{\prime}}^2},
\label{Ft}
\ee

\noindent where $f$ is the statistical rate function, $t$ the partial
half-life for the transition, $\delta_R$ is the calculated nucleus-dependent
radiative correction, $\delta_C$ the calculated isospin-breaking correction,
and $K$ is a known \cite{Ha90} constant.  The effective coupling
constant relates to the primitive one via
$\GV^{\prime} = \GV (1 + \DRV )^{1/2}$,
where $\DRV$ is a calculated nucleus-independent radiative
correction.  For tests of the CVC hypothesis it is not necessary to
consider this correction.

In this report, we highlight two recent experiments
\cite{Sa95,Ha94} at Chalk River aimed at shedding light on the two
corrections in Eq.\,(\ref{Ft}).  By far, the largest contribution
to the $\F t$-value uncertainty comes from the calculation of
$\delta_R$ and $\delta_C$, not from the experimental data.
There has been a suggestion that the calculated corrections, to date,
are not complete since the data seem to display a small residual
$Z$-dependence in the $\F t$ values.
In this case, the best overall $\F t$ value might be
taken from a curve fitted to the individual data and extrapolated
\cite{Wi90} to $Z=0$.  In such fits the deduced $V_{ud}$ matrix
element is larger and the unitarity test on the CKM matrix satisfied.

In order to establish whether such phenomenological fitting can be
justified, two experiments were mounted at Chalk River: (1) to provide
a new data point closer to $Z=0$ by measuring the branching ratio for
the lightest superallowed Fermi beta emitter, $^{10}$C, and (2) to
measure branching ratios to non-analogue $0^{+}$ states in the daughter
nucleus, which tests the model predictions for isospin-mixing
corrections.

\section{The $^{10}$C branching-ratio experiment} \label{Cbr}

The decay of $^{10}$C takes place mainly through a strong Gamow-Teller
transition to an excited $1^{+}$ state at 718 keV in $^{10}$B, while
only about 1.5\% of the decays go to the isobaric analogue $0^{+}$
state at 1740 keV.  The superallowed branching ratio is simply given
by the ratio of the number of gamma rays at 1022 (1740 $-$ 718) keV
to that at 718 keV, \ie

\be
B(0^{+} \rightarrow 0^{+}) = \frac{R(1022)}{R(718)} =
\frac{Y(1022)}{Y(718)} \frac{\epsilon (718)}{\epsilon (1022)} ,
\label{ratio}
\ee

\noindent with $R$ being the emission rate, $Y$ the observed yield, and
$\epsilon$ the detection efficiency at a given $\gamma$-ray energy.
Any measurement, however,
requires excellent statistics to yield precision of a few parts per thousand
on such a weak branch.  In addition, since the isobaric analogue state
populated by the superallowed branch is deexcited by the emission of a
1022 keV gamma ray, it is necessary to minimize and account for the
pileup of 511 keV annihilation radiation that disturbs the measurement.

The experiment was therefore performed on a large gamma-ray array:
the $8 \pi$ spectrometer at Chalk River.  Here the total detector
efficiency is shared by the $N$ independent detectors, and the
pileup-to-signal
ratio is decreased by a factor of $N$.  The $8 \pi$ spectrometer
is composed of 20 Compton-suppressed 25\% HPGe detectors surrounding a
72-element BGO inner ball.  In addition to the twentyfold reduction
in the 511 pileup signal obtained because of the geometry of the array
itself, a further reduction is obtained via the pileup rejection
system on each germanium detector, which has a mean resolving time
of roughly 420 ns.

The experiment comprised two interleaved measurements.
One, the relative $\gamma$-ray
yield measurement, was a repeated cycle in which the activity
was first produced by a $(p,n)$ reaction on a $^{10}$B target mounted
in the centre
of the $8 \pi$ spectrometer; then the beam was turned off and the
$\beta$-delayed gamma rays from the decay of $^{10}$C observed in
singles mode.  The second measurement, that of the relative gamma-ray
efficiency, was performed in beam with $\gamma$-$\gamma$ coincidences
recorded from the deexcitation of the 2154 keV level in $^{10}$B,
which was populated by the $(p,p^{\prime})$ reaction.  Further
details are given in ref. \cite{Sa95} \nl .

After a number of corrections were applied, the total branch to the
isobaric analogue state was determined to be

\be
B(0^{+} \rightarrow 0^{+}) = [ 1.4625 \pm 0.0020 ({\rm stat})
\pm 0.0015 ({\rm syst})] \% ,
\label{BR}
\ee

\noindent where the systematic uncertainty is the one attributed to
the sum of all experimental corrections; it should be added quadratically to
the statistical uncertainty.  This result agrees with, but is
substantially more precise than, previous measurements:  $(1.465 \pm
0.014)\%$, ref.\cite{Ro72} \nl ; $(1.473 \pm 0.007)\%$, ref.\cite{Na91}
\nl ; $(1.465 \pm 0.009)\%$, ref.\cite{Kr91} \nl .  When
our results are averaged with the previous measurements
and combined with $Q_{EC} = 1907.77(9)$ keV \cite{Ba89}
and $t_{1/2} = 19.209(12)$ s \cite{Az74}\nl , corrected for electron
capture (0.296\%), the $ft$ value obtained is

\be
ft(^{10}{\rm C}) = 3040.1 \pm 5.1 s.
\label{10Cft}
\ee

\section{Branching ratios to non-analogue $0^{+}$ states} \label{nonbr}

The charge-dependent correction $\delta_C$, introduced in Eq.\,(\ref{Ft}),
reflects differences between the initial- and final-state wavefunctions,
and thus is strongly nuclear-structure dependent.  The two most
complete
calculations of this correction by Towner-Hardy-Harvey (THH)
\cite{THH77,To89} and Ormand-Brown (OB) \cite{OB89,OB95} show
qualitative agreement in that the large variations in $\delta_C$ from
nucleus to nucleus are similar in both models.  However, the models
differ in their values for the absolute magnitude of the correction.

Both calculations identify two separate contributions to the
charge-dependent
correction.  The larger, radial-overlap part, $\delta_{RO}$,
arises from the fact that protons are less bound than neutrons,
so the (initial)
proton wavefunction imperfectly overlaps the (final) neutron one.
The smaller, isospin-mixing part $\delta_{IM}$ results from different
degrees of configuration mixing in the wavefunctions of members of
an isospin multiplet.  This latter correction is amenable to
experimental test.  If we denote the Fermi
matrix element for the ground-state transition as
$\langle M_0 \rangle$ and that
for the non-analogue transition to an excited $0^{+}$ state as
$\langle M_1 \rangle $, then,
for states with $(J^{\pi},T) = (0^{+},1)$,

\bea
\langle M_0 \rangle ^2 & = & 2 (1 - \delta_C ) \simeq 2 (1 - \delta_{IM})
(1 - \delta_{RO})
\label{M0}  \\
\langle M_1 \rangle ^2 & = & 2 \delta_{IM}^1 (1 - \delta_{RO}),
\label{M1}
\eea

\noindent where $\delta_{IM}^1$ is essentially the admixture of the
$0^{+}$ ground state into the first excited $0^{+}$ state.  The branching
ratio $B_1$ to the latter is

\be
B_1 \approx \frac{t_0}{t_1} = \frac{f_1}{f_0} \frac{f_0 t_0}{f_1 t_1}
= \frac{f_1}{f_0} \frac{2 \delta_{IM}^1}{2 (1-\delta_{IM})}
\approx \frac{f_1}{f_0} \delta_{IM}^1 ,
\label{B1}
\ee

\noindent where subscripts 0 and 1 again indicate the ground state and
excited $0^{+}$ state, respectively.

Experiments \cite{Ha94} at Chalk River have searched for non-analogue
$0^{+} \rightarrow 0^{+}$ transitions in the decays of four
superallowed beta emitters, \viz : $^{38m}$K, $^{46}$V, $^{50}$Mn
and $^{54}$Co.  Samples of $^{38m}$K were produced with $(\alpha,n)$
reactions and of the other three emitters with $(p,n)$  reactions,
all on isotopically enriched targets.  The experiments were performed at
the TASCC facility, with a helium-jet gas-transfer system used to
convey activities from the target chamber to a low-background counting
location.

A 68\% HPGe detector was used to look for the characteristic $\beta$-delayed
gamma rays from excited $0^{+}$ states in the daughter.  Since the
non-analogue transitions are very weak (ppm level) strong samples
(MBq level) were required.  Passive shielding installed in front of
the detector crystal prevented direct exposure to the high flux of
energetic positrons from the dominant ground-state branch, but the
resulting bremsstrahlung radiation was intense enough to obscure any
weak gamma-ray branches.  To overcome this limitation, two thin
plastic scintillators were positioned in front of the HPGe detector
and on either side of the collected sample.  All recorded gamma rays were
tagged with the status of the positron events in the scintillators.
With the scintillator information invoked, the level of continuous
background in the HPGe spectrum was reduced by a factor of 400
compared to the singles result.  This permitted the observation of
$\beta$-delayed gamma-ray branches down to the 10 ppm level.

\begin{table}[t]
\begin{center}
\caption{Analogue-symmetry-breaking corrections \label{imc}}
\vskip 1mm
\begin{tabular}{llllcll}
\hline \\[-3mm]
 & Expt.(\%) & \multicolumn{5}{c}{Theory(\%)} \\
 & & \multicolumn{2}{c}{THH} & & \multicolumn{2}{c}{OB} \\
\cline{3-4} \cline{6-7}  \\[-3mm]
Nuclide & $~~~\delta_{IM}^1$ & $~~~\delta_{IM}^1$ & $~~~\delta_{IM}$
 &  & $~~~\delta_{IM}^1$ & $~~~\delta_{IM}$ \\
\hline \\[-2mm]
$^{38m}$K & $ < 0.28 $ & 0.096(2) & 0.100(2) & & & \\
$^{46}$V & 0.053(5) & 0.046(5) & 0.087(10) & & 0.054(50) & 0.094(50) \\
$^{50}$Mn & $ < 0.016 $ &  0.051(23) & 0.068(30) & & 0.015(50) & 0.017(50) \\
$^{54}$Co & 0.035(5) & 0.037(8) & 0.045(5) & & 0.003(50) & 0.006(50) \\
\hline
\end{tabular}
\end{center}
\end{table}

The results obtained on the four superallowed $\beta$ emitters are given
in Table \ref{imc}, together with theoretical values computed by THH
\cite{To89}
and updated in \cite{Ha94} \nl , and by OB \cite{OB95} \nl .
The THH calculations are in good agreement with experiment for $^{46}$V
and $^{54}$Co but overestimate slightly for $^{50}$Mn.  OB were able to
reproduce the $^{46}$V result and the $^{50}$Mn limit, but these
authors had difficulty with their shell-model calculations for
$^{54}$Co, which they discuss in their manuscript \cite{OB95} \nl .
The authors assigned a large error estimate to all their $\delta_{IM}$
computations.
On balance, the calculations are in reasonable agreement with
experiment showing
that this part of the isospin-symmetry-breaking correction is under
control.

\section{Current status} \label{cs}

World data on $Q$-values, lifetimes and branching ratios were
thoroughly surveyed \cite{Ha90} in 1989 and updated again \cite{TH95} this
year.  The $ft$-values and $\delta_R$ correction
are taken from this latter reference.  Ormand and Brown \cite{OB95}
have just released a revised calculation of $\delta_C$ but, as
with their earlier calculation \cite{OB89} \nl , there is a systematic
difference from that of THH \cite{THH77} \nl ; however the magnitude
of this difference is reduced by about a factor of two.  This
difference represents a `systematic' uncertainty of $\pm 0.04 \%$
that is put to one side for the test of CVC, but must be
applied at a later stage to the average $\F t$ value.  We adopt for
$\delta_C$ the unweighted average of the THH and OB values.

\begin{figure}[t]
\begin{center}
\setlength{\unitlength}{0.240900pt}
\ifx\plotpoint\undefined\newsavebox{\plotpoint}\fi
\sbox{\plotpoint}{\rule[-0.175pt]{0.350pt}{0.350pt}}%
\begin{picture}(1500,900)(0,0)
\tenrm
\sbox{\plotpoint}{\rule[-0.175pt]{0.350pt}{0.350pt}}%
\put(264,315){\rule[-0.175pt]{4.818pt}{0.350pt}}
\put(242,315){\makebox(0,0)[r]{3070}}
\put(1416,315){\rule[-0.175pt]{4.818pt}{0.350pt}}
\put(264,473){\rule[-0.175pt]{4.818pt}{0.350pt}}
\put(242,473){\makebox(0,0)[r]{3075}}
\put(1416,473){\rule[-0.175pt]{4.818pt}{0.350pt}}
\put(264,630){\rule[-0.175pt]{4.818pt}{0.350pt}}
\put(242,630){\makebox(0,0)[r]{3080}}
\put(1416,630){\rule[-0.175pt]{4.818pt}{0.350pt}}
\put(354,158){\rule[-0.175pt]{0.350pt}{4.818pt}}
\put(354,113){\makebox(0,0){5}}
\put(354,767){\rule[-0.175pt]{0.350pt}{4.818pt}}
\put(580,158){\rule[-0.175pt]{0.350pt}{4.818pt}}
\put(580,113){\makebox(0,0){10}}
\put(580,767){\rule[-0.175pt]{0.350pt}{4.818pt}}
\put(805,158){\rule[-0.175pt]{0.350pt}{4.818pt}}
\put(805,113){\makebox(0,0){15}}
\put(805,767){\rule[-0.175pt]{0.350pt}{4.818pt}}
\put(1030,158){\rule[-0.175pt]{0.350pt}{4.818pt}}
\put(1030,113){\makebox(0,0){20}}
\put(1030,767){\rule[-0.175pt]{0.350pt}{4.818pt}}
\put(1256,158){\rule[-0.175pt]{0.350pt}{4.818pt}}
\put(1256,113){\makebox(0,0){25}}
\put(1256,767){\rule[-0.175pt]{0.350pt}{4.818pt}}
\put(264,158){\rule[-0.175pt]{282.335pt}{0.350pt}}
\put(1436,158){\rule[-0.175pt]{0.350pt}{151.526pt}}
\put(264,787){\rule[-0.175pt]{282.335pt}{0.350pt}}
\put(45,562){\makebox(0,0)[l]{\shortstack{$\F t(s)$}}}
\put(850,23){\makebox(0,0){$Z$ of daughter}}
\put(264,158){\rule[-0.175pt]{0.350pt}{151.526pt}}
\put(354,454){\circle*{24}}
\put(444,306){\circle*{24}}
\put(670,315){\circle*{24}}
\put(850,318){\circle*{24}}
\put(940,296){\circle*{24}}
\put(1030,545){\circle*{24}}
\put(1120,454){\circle*{24}}
\put(1211,435){\circle*{24}}
\put(1301,384){\circle*{24}}
\put(354,284){\rule[-0.175pt]{0.350pt}{81.665pt}}
\put(344,284){\rule[-0.175pt]{4.818pt}{0.350pt}}
\put(344,623){\rule[-0.175pt]{4.818pt}{0.350pt}}
\put(444,224){\rule[-0.175pt]{0.350pt}{39.508pt}}
\put(434,224){\rule[-0.175pt]{4.818pt}{0.350pt}}
\put(434,388){\rule[-0.175pt]{4.818pt}{0.350pt}}
\put(670,249){\rule[-0.175pt]{0.350pt}{31.799pt}}
\put(660,249){\rule[-0.175pt]{4.818pt}{0.350pt}}
\put(660,381){\rule[-0.175pt]{4.818pt}{0.350pt}}
\put(850,243){\rule[-0.175pt]{0.350pt}{36.376pt}}
\put(840,243){\rule[-0.175pt]{4.818pt}{0.350pt}}
\put(840,394){\rule[-0.175pt]{4.818pt}{0.350pt}}
\put(940,199){\rule[-0.175pt]{0.350pt}{46.975pt}}
\put(930,199){\rule[-0.175pt]{4.818pt}{0.350pt}}
\put(930,394){\rule[-0.175pt]{4.818pt}{0.350pt}}
\put(1030,473){\rule[-0.175pt]{0.350pt}{34.690pt}}
\put(1020,473){\rule[-0.175pt]{4.818pt}{0.350pt}}
\put(1020,617){\rule[-0.175pt]{4.818pt}{0.350pt}}
\put(1120,369){\rule[-0.175pt]{0.350pt}{40.953pt}}
\put(1110,369){\rule[-0.175pt]{4.818pt}{0.350pt}}
\put(1110,539){\rule[-0.175pt]{4.818pt}{0.350pt}}
\put(1211,350){\rule[-0.175pt]{0.350pt}{40.953pt}}
\put(1201,350){\rule[-0.175pt]{4.818pt}{0.350pt}}
\put(1201,520){\rule[-0.175pt]{4.818pt}{0.350pt}}
\put(1301,300){\rule[-0.175pt]{0.350pt}{40.712pt}}
\put(1291,300){\rule[-0.175pt]{4.818pt}{0.350pt}}
\put(1291,469){\rule[-0.175pt]{4.818pt}{0.350pt}}
\sbox{\plotpoint}{\rule[-0.350pt]{0.700pt}{0.700pt}}%
\put(264,388){\usebox{\plotpoint}}
\put(264,388){\rule[-0.350pt]{282.335pt}{0.700pt}}
\end{picture}
\vskip 1mm
\caption{$\F t$ values for the nine precision data and the best
least-squares one-parameter fit  \label{fig1}}
\end{center}
\end{figure}
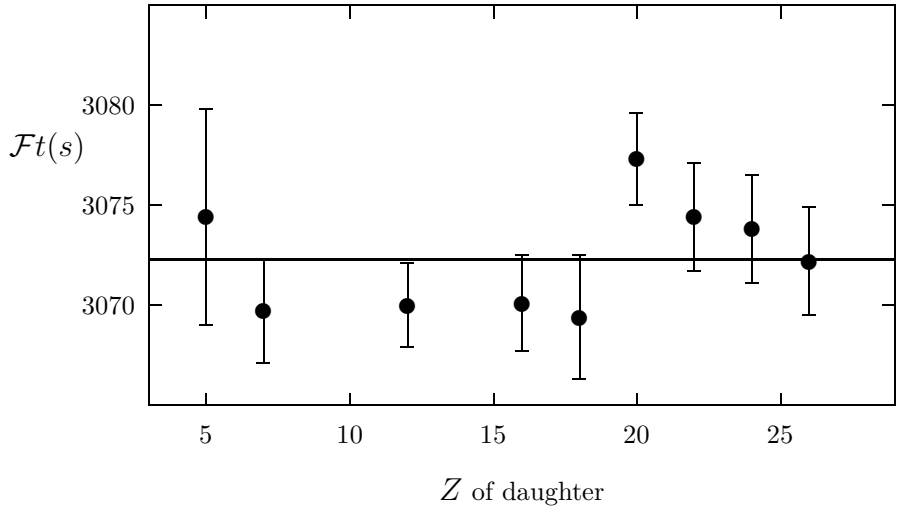

The results for the nine $\F t$ values are displayed in Fig. \ref{fig1}.
The uncertainties shown reflect the experimental uncertainties and an
estimate of the {\em relative} uncertainties in $\delta_C$.  There is
no statistically significant evidence of inconsistencies in the
data ($\chi^2/\nu = 1.2$),
thus verifying the expectation of CVC at the level of
$4 \times 10^{-4}$, the fractional uncertainty quoted on the
average $\F t$ value ($3072.3 \pm 1.0$ s).  In using the average
$\F t$ value to determine $V_{ud}$ and test CKM unitarity it is
important to incorporate the `systematic' uncertainty in $\delta_C$
just referred to.  The result is

\be
\F t = 3072.3 \pm 2.0 ~{\rm s}.
\label{avgFt}
\ee

\noindent With this value, an estimate \cite{Si94} of the
nucleus-independent radiative correction of $\DRV = (2.40 \pm 0.08)\%$,
and the weak vector coupling constant \cite{PDG94} derived from
muon decay,
we obtain

\be
V_{ud} = 0.9740 \pm 0.0005 .
\label{Vud}
\ee

\noindent The quoted uncertainty is dominated by uncertainties
in the theoretical
corrections, $\DRV$ and $\delta_C$.  On adopting the values \cite{PDG94}
of $V_{us}$
and $V_{ub}$ from the Particle Data Group, the sum
of squares of the elements in the first row of the CKM matrix,

\be
\mids V_{ud} \mids^2 +
\mids V_{us} \mids^2 +
\mids V_{ub} \mids^2 = 0.9972 \pm 0.0013,
\label{unit}
\ee

\noindent differs from unity at the $98\%$ confidence level.

The significance of this apparent non-unitarity is not yet settled.
It may, of course, indicate the need for some extension to the
three-generation Standard Model -- perhaps in the form of right-hand
currents or additional gauge bosons.  However, it may also reflect
some undiagnosed inadequacy in the evaluation of $V_{us}$
(as already suggested for other reasons in ref.\cite{Ga92} ) or
possibly in the $\delta_C$ corrections used to determine $V_{ud}$.
There is little scope left by the data in Fig.\ref{fig1} for
introducing significant additional $Z$-dependence in $\delta_C$,
as has been suggested recently \cite{Wi90} \nl .

\end{document}